# Systematic evolution of superconducting pairing strength and Seebeck coefficients in correlated infinite-layer $La_{1-x}Sr_xNiO_2$

Motoki Osada,[1,2]* Shusaku Imajo,[3] Yuji Seki,[4] Kousuke Ishida,[1] Tsutomu Nojima,[1] Kohei Fujiwara,[1,5] Koichi Kindo,[3] Yusuke Nomura,[1,4,6] Atsushi Tsukazaki[1,2]

[1]Institute for Materials Research, Tohoku University, Sendai, Miyagi 980-8577, Japan.

[2]Quantum-Phase Electronics Center (QPEC) and Department of Applied Physics, The University of Tokyo, Hongo, Tokyo 113-8656, Japan.

[3]The Institute for Solid State Physics, The University of Tokyo, Kashiwa, Chiba 277-8581, Japan.

[4]Department of Applied Physics and Physico-Informatics, Keio University, Yokohama, Kanagawa 223-8522, Japan.

[5]Department of Chemistry, Rikkyo University, Toshima, Tokyo 171-8501, Japan.

[6]Advanced Institute for Materials Research (WPI-AIMR), Tohoku University, Sendai, Miyagi, Japan.

*Email: osada@ap.t.u-tokyo.ac.jp

**Abstract**

The recently discovered superconducting infinite-layer nickelates offer a novel platform to explore an exotic pairing mechanism in multi-band systems towards high-temperature superconductivity and associated rich quantum phases, contrasting with cuprates. Here, we show that infinite-layer $(La,Sr)NiO_2$ exhibits strong-coupling superconductivity, resilient to in-plane magnetic fields exceeding 47 T at optimal doping—more than twice the Pauli limit for conventional BCS superconductors. This violation becomes pronounced towards the underdoped regime, implying an intriguing evolution of pairing glue. The unexpected observation of positive Seebeck coefficients in this regime indicates the presence of nontrivial electron correlations. Furthermore, our comprehensive investigation across the superconducting dome reveals that the evolution of (thermo)electric normal-state properties—specifically, the sign changes of the Hall and Seebeck coefficients—coincide with the evolution of superconducting anisotropy and pairing strength. This demonstrates a definitive link between electron correlations and strong-coupling superconductivity in $(La,Sr)NiO_2$, contributing to a unified framework for understanding unconventional superconductivity.

**Introduction**

Unconventional superconductivity frequently emerges in close proximity to strongly correlated phases, suggesting that a thorough understanding of these correlations is crucial for uncovering the underlying mechanisms of high-temperature superconductivity (*1*, *2*). In copper oxide superconductors, the strong electron correlation is particularly pronounced in the underdoped regime, where it yields a diverse array of quantum phases, including antiferromagnetic insulators, pseudogap states, and charge orders (*3*). Strong-coupling superconductivity, characterized by electron pairs bound by robust interactions, represents a large superconducting energy gap to break the electron pairs. Such strong interactions, which can also involve spin density wave or charge density wave, can result in a higher superconducting transition temperature $T_c$ than those predicted by conventional Bardeen-Cooper-Schrieffer (BCS) theory (*4*). The mechanism for the strong pairing glue and its relation to electron correlations in unconventional superconductivity has been a subject of extensive debate over the past few decades.

The discovery of superconductivity in infinite-layer nickelate thin films has provided a unique opportunity to address these issues from a new perspective (*5*). The family of nickelates possesses a layered crystal structure and a $3d^9$ electronic configuration, reminiscent of copper oxide superconductors (*6–11*). The normal states of infinite-layer nickelates, (Nd,Sr)NiO$_2$, exhibit a linear-in-temperature (*T*-linear) resistivity at optimal doping—a hallmark of the strange metal behavior observed in many unconventional superconductors (*12–15*). However, key distinctions evoking to relatively weak electron correlations remain, such as the absence of antiferromagnetic long-range order (Mott insulator), possibly due to the multi-orbital electronic nature involving Ni $3d$ and rare-earth $5d$ orbitals (*16–18*). Recently, spectroscopic observations of charge order in infinite-layer nickelates have been reported, although this remains a subject of ongoing debate (*19–22*). To date, superconducting infinite-layer nickelates have been realized only in thin-film form through a topotactic transition from the perovskite phase to the highly metastable infinite-layer phase (Fig. 1A). The difficulty in obtaining a clean surface via this topotactic reaction has limited the experimental determination on the doping evolution of the band structure using the spectroscopic techniques. Given this situation, a multifaceted and comprehensive investigation of the properties of the normal and superconducting states across the superconducting dome of La$_{1-}$

$_x$Sr$_x$NiO$_2$ through the transport measurements could provide a crucial insight into fundamental understanding of the electronic landscape and unconventional superconductivity in this system.

In this study, we prepared the La$_{1-x}$Sr$_x$NiO$_2$ films on SrTiO$_3$ substrates by pulsed-laser deposition and ex-situ low-temperature annealing with CaH$_2$ powder (see Fig. S1 and S2 for sample characterization). As shown in the high-angle annular dark-field (HAADF) scanning transmission electron microscopy (STEM) cross-sectional image of La$_{0.84}$Sr$_{0.16}$NiO$_2$ film (Fig. 1B), we have successfully obtained high-quality film with abrupt interface and minimized disorder. To identify the optimal doping level $x$, we compare the temperature dependence of resistivity $\rho_{xx}$ with representative curves for $x$ = 0.08, 0.16, and 0.24 shown in Fig. 1C. We observed weakly insulating behavior at $x$ = 0.08, a sharp superconducting transition at $x$ = 0.16, and a suppressed $T_c$ at $x$ = 0.24. Based on these results, the onset $T_c$ was plotted as a function of $x$, as shown in Fig. 1D. The resulting dome-shaped trend highlights an optimal doping region centered around $x$ = 0.16 (white region) with the underdoped (red) and overdoped (blue) regions identified accordingly. The onset of $T_c$ values for La$_{1-x}$Sr$_x$NiO$_2$ films in this study (blue solid circles) are comparable to, or exceed to, those reported in previous studies (gray and green open diamonds) (*23*, *24*), confirming the high quality of our samples. For optimal $x$ = 0.16, high-field magnetoresistance measurements reveals that superconductivity survives at approximately 47 T at 1.6 K (90% of normal state resistance as a criterion) under the in-plane magnetic field, as shown in the magnetoresistance curves (Fig. 1E). This large $H_{c2}^{ab}$ in the optimal doping is comparable to previously reported values for La$_{0.8}$Sr$_{0.2}$NiO$_2$ and is nearly twice as large as those of Nd$_{0.825}$Sr$_{0.175}$NiO$_2$ in prior literatures (*25–29*). While the intrinsic behavior of $H_{c2}$ is inferred to be largely similar across the infinite-layer nickelates, notable differences in the magnitude of the superconducting upper critical field across rare-earth elements have been attributed to the 4$f$ electrons of rare-earth ions (*29*).

Assuming a negligible contribution of orbital pair-breaking effect under the in-plane magnetic field, $H_{c2}^{ab}$ serves as an approximate measure of the Pauli-limiting field. In Fig. 1F, we plot $\mu_0 H_{c2}^{ab}/T_c$ as a function of $T_c$ to assess pairing strength relative to other materials (*30–44*). Firstly, we should point out that $\mu_0 H_{c2}^{ab}/T_c$ = 1.86 T/K (Pauli limit: broken line) is expected as a maximum value in the weak-coupling BCS superconductors. In contrast, $\mu_0 H_{c2}^{ab}/T_c$ values exceeding 1.86 have been reported in several materials, offering valuable insights into the mechanisms driving the enhancement of $H_{c2}/T_c$. For example, the Ising-type spin valley locking owing to large spin-orbit coupling, which originates from the asymmetric trigonal crystal structure, enables the large value of MoS$_2$ monolayer much over the Pauli limit (*35*). Other prominent examples include heavy fermion systems (UPt$_3$ and CeCoIn$_5$) (*31*, *34*), magic-angle twisted trilayer graphene (MATTG) (*43*), organic compound (κ-(BEDT-TTF)$_4$Hg$_{2.89}$Br$_8$) (*44*), iron-based superconductors (*36*, *38*, *42*), and copper oxides (*40*, *41*). These varieties of the coupling strength reflect the different feature of superconducting gap of each material. Among them, the value of La$_{1-x}$Sr$_x$NiO$_2$ films appears to be definitively high, suggesting the presence of strong electron pairing glue.

**Results and Discussion**

**Pairing strength**

We further investigated the in-plane $H_{c2}^{ab}$ and out-of-plane upper critical fields $H_{c2}^{c}$ of superconducting samples, $x = 0.12$ (Fig. 2A for $H_{c2}^{ab}$) and 0.24 (Fig. 2B for $H_{c2}^{ab}$), under high-magnetic fields (see Fig. S3 and S4 for all magnetoresistance curves, and Fig. S5 and S6 for all $H_{c2}$–$T$ phase diagrams in-plane and out-of-plane, respectively). Here, we define $H_{c2}^{ab(c)}$ as the magnetic field with 90% of the normal resistance. At the lowest measurement temperature of 1.5 K, the $\mu_0 H_{c2}^{ab}$ of underdoped film is close to 26 T (blue curve in Fig. 2A), which is clearly higher than that of overdoped films of nearly 19 T (blue curve in Fig. 2B). This comparison, presented in Fig. 2C, shows that the underdoped sample has markedly enhanced resilience to in-plane magnetic fields, despite both samples having similar $T_c$ values of around 6 K. Even when we employed 50% of the normal-state resistance as a criterion and constructed an $H_{c2}$–$T$ phase diagram, $H_{c2}$ exhibits a similar behavior (Fig. S7). The pronounced difference in $H_{c2}^{ab}$ suggests a contrasting the pairing strength between the two regimes. Importantly, our results show a systematic decrease in $H_{c2}(0)/T_c$ with doping, which is inconsistent with disorder effects such as orbital pair-breaking or impurity scattering. Moreover, the experimentally determined $H_{c2}(0)$ values are significantly lower than the orbital-limited field estimated from the initial slope of $H_{c2}(0)$ near $T_c$ (the broken line in Fig. S5), providing empirical evidence that the upper critical field is governed by the Pauli limit. These findings strongly suggest that the variation of $H_{c2}(0)/T_c$ reflects an intrinsic suppression of the pairing interaction rather than disorder effects. Recently, an isotropic quantum Griffiths singularity (QGS) in Nd$_{0.8}$Sr$_{0.2}$NiO$_2$ thin films, associated with Kondo scattering, has been reported, suggesting a possible connection to exotic superconductivity (*45*). In the underdoped regime (Sr *x* =0.12 and 0.14), we observed a crossing of resistance curves measured at different temperatures

(Fig. S3 and S4). The potential link between quantum fluctuations associated with QGS and strong correlations in La$_{1-x}$Sr$_x$NiO$_2$ requires further investigation.

Figure 2D shows the superconducting anisotropy using $\gamma_H = H_{c2}^{ab}/H_{c2}^c$ as a function of $x$. While the optimal and overdoped regimes exhibit nearly isotropic behavior ($\gamma_H \sim 1$), a sharp enhancement of $\gamma_H$ in the underdoped regime suggests a more anisotropic superconducting state. We further evaluated the pairing strength by plotting $\mu_0 H_{c2}^{ab}/T_c$ in Fig. 2E, which indicates the enhancement in underdoped regime. Similarly, as shown in Fig. S8, the value of $\mu_0 H_{Pauli}^{ab}(0)/T_c$ also increases in this regime. Since $H_{Pauli}(0)$ is closely related to the magnitude of the zero-temperature superconducting gap $\Delta_0$ via the relation $\mu_0 H_{Pauli}(0)/T_c \propto \Delta_0/T_c$, these observations suggest that pairing interactions are enhanced in the underdoped regime (see Table S1 for orbital- and Pauli-limited fields and coherence lengths for all doping levels). The consistently large $\mu_0 H_{c2}^{ab}/T_c$ values observed across the entire doping range—exceeding the weak-coupling limit of 1.86 T/K and reaching as high as 4.4 T/K at underdoped regime ($x = 0.12$)—demonstrates the strong-coupling superconductivity in La$_{1-x}$Sr$_x$NiO$_2$. Interestingly, the increase in both anisotropy (Fig. 2D) and pairing strength (Fig. 2E) coincides in the underdoped regime. As shown in Fig. S9, we further performed electrical transport measurements on the underdoped region (Sr $x = 0.08$) under high-magnetic fields ($B_{\parallel c} = 25$ T) and found that the resistive upturn remains nearly unchanged under high-magnetic fields. This suggests that the effects of disorder or localization in the underdoped regime on transport behavior is small. The emergence of anisotropic strong-coupling superconductivity indicates a major role of Ni $d_{x^2-y^2}$ orbital, which can host strong electron correlations.

**DFT calculations and (thermo)electric transport**

To address possible electron correlation effects, we investigate the normal-state properties by conducting electronic band calculations based on density functional theory (Fig. 3A–D) and (thermo)electric transport measurements for La$_{1-x}$Sr$_x$NiO$_2$ films (Fig. 3E–H)—namely, the Hall and Seebeck coefficients, which are sensitive to charge carrier density and particle-hole asymmetry, respectively, and together offer a comprehensive perspective on the electronic structure. The calculated band structure is displayed with three colors corresponding to the dominant band characters of Ni $3d_{x^2-y^2}$ (green in Fig. 3B), interstitial $s$ (blue shown in Fig. 3C), and La $5d_{3z^2-r^2}$ orbitals (orange). Fermi surface is mainly composed of the bands with the interstitial $s$ and Ni $3d_{x^2-y^2}$ characters. Comparing the electronic bands of underdoped (Fig. 3A) and the overdoped regimes (Fig. 3D), we observe an energy shift in the electron pockets of the La $5d_{3z^2-r^2}$ and interstitial $s$ bands, yet the Ni $3d_{x^2-y^2}$ band remains dominant at the Fermi surface. Based on these DFT-calculated bands, we theoretically estimate the Hall coefficient $R_H$ and Seebeck coefficient $S_{xx}$ by accounting for the basic electron correlation effects, i.e., mass enhancement and temperature-dependent scattering rates (see Supplementary Materials for detailed description). These theoretically estimated values are then compared with experimental results of underdoped ($x = 0.08$) and overdoped ($x = 0.28$) regimes as a function of temperature (Fig. 3E–H, see also Fig. S10 and S11 for all Hall and Seebeck measurements).

The calculated transport coefficients exhibit minimal variation between the two regimes because the DFT band structure, especially the dominant Ni $3d_{x^2-y^2}$ band, remains unchanged significantly. However, intriguingly, the experimental transport properties show a substantial variation. The experimental data of the Hall coefficient $R_H$ exhibit typical behaviors: fully negative across all temperature ranges for the underdoped $x = 0.08$ (Fig. 3E) and a clear sign reversal around 60 K for the overdoped $x = 0.28$ (Fig. 3G), consistent with previous studies (*23*). Additionally, this study provides experimental data on the Seebeck coefficient, revealing a sign change of $S_{xx}$ between two doping regimes (Fig. 3F, H).

In the overdoped regime, we find good agreement between the experimental results and the DFT-based calculations. Since the Hall and Seebeck coefficients are sensitive to carrier density and particle-hole asymmetry near Fermi level, respectively, and provide an insight into the electronic structure from different perspectives, the consistent agreement in *both* quantities offers compelling evidence that the electronic structure in the overdoped regime fit well to Fermi-liquid-like description.

In contrast, we find discrepancies between experimental data and calculations in the underdoped regime. In particular, the positive experimental $S_{xx}$ values below 200 K are inconsistent with the negative values predicted by DFT-based calculations (Fig. 3F). According to the Mott formula, the Seebeck coefficient is a measure for the energy-derivative of density-of-states or the scattering rate. We systematically compared our experimental results with DFT band calculations, incorporating mass renormalization and temperature-dependent scattering effects. While these factors can explain the overdoped behavior, they fail to account for the positive $S_{xx}$ observed in the underdoped regime. Since mass renormalization changes the magnitude but not the sign of the carrier velocity, the presence of positive $S_{xx}$ requires further consideration beyond the conventional DFT-based picture, indicating unexpected particle-hole asymmetry in band dispersion and/or scattering rate. Such a drastic modulation of the electronic structure could be caused by band folding due to symmetry breaking associated with the phase transition, or by band reconstruction and anomalous particle-hole asymmetric scattering rate driven by the proximity to the Mott insulating phase (*45–47*). For example, the theoretical simulations of the doped Hubbard model shows that strong electron correlations can invert the sign of Seebeck coefficient from negative to positive (*48*). This indicates that, although the Mott insulating phase is absent in the infinite-layer nickelates due to the self-doping effect, the Ni $3d_{x^2-y^2}$ band may behave like a doped Mott insulator. Similar to other strongly correlated superconductors, such as copper oxides, iron-pnictides, and heavy fermion compounds, many exotic phases may be hidden in the underdoped regime in infinite-layer nickelates.

**Consistency of normal state and superconducting state**

The distinct contrast between the underdoped and overdoped regimes is clearly manifested in both the (thermo)electric transport properties in the normal state (Fig. 3E–H) and the anisotropy and pairing strength at superconducting state (Fig. 2D, E). We map out $R_H$ (Fig. 4A), $S_{xx}/T$ (Fig. 4B), the onset $T_c$ (Fig. 4C), and $\mu_0 H_{c2}^{ab}/T_c$ (Fig. 4D) as functions of temperature and doping concentration $x$. The color (contour) maps in all figures visualize the key threshold. The whitish region with the broken line represents the distinguishable criteria: the negative (blue) or positive (red) sign in Fig. 4A and B, and 1.86 T/K for the weak-coupling limit in Fig. 4D. Across the boundary around $x_c \sim 0.16$, the sign reversal is consistently revealed in $R_H$ and $S_{xx}/T$, indicating dramatic change in the electronic structure owing to increase of the $x$ content. In particular, the sign reversal of $S_{xx}/T$ in La$_{1-x}$Sr$_x$NiO$_2$ across the critical doping concentration $x_c \sim 0.16$ (all data is given in Fig. S12) cannot be explained by conventional mass renormalization or simple scattering effects, indicating the presence of nontrivial electron correlations in the underdoped regime. Notably, the evolution of $S_{xx}/T$ is also discussed in YBa$_2$Cu$_3$O$_y$, La$_{1.8-x}$Eu$_{0.2}$Sr$_x$CuO$_4$, and La$_{1.6-x}$Nd$_{0.4}$Sr$_x$CuO$_4$, associated with reconstruction of band structures (*46*, *47*). This boundary concentration $x_c$ is located around the center of the superconducting dome (Fig. 4C), suggesting that an evolution of the electronic state would be related to the highest $T_c$. Finally, the pairing strength evaluated by $\mu_0 H_{c2}^{ab}/T_c$ surprisingly exhibits a sudden enhancement in underdoped regime below $x_c$ (deep red in Fig. 4D), which could be linked to the reconstruction of electronic structure driven by electron correlation effect. The electron correlation at the underdoped regime will be the focus of further discussion as like copper oxide superconductors.

In summary, comprehensive experimental and theoretical studies on $La_{1-x}Sr_xNiO_2$ film has been carried out as a function of systematic Sr content $x$. Our findings point towards a strong-coupling mechanism for nickelate superconductivity that is deeply tied to electron interactions, particularly in the underdoped regime. While the electron correlation effect in nickelates has been thought to be weak, we found that the correlation effect in the underdoped $La_{1-x}Sr_xNiO_2$ is of sufficient strength to induce band reconstruction, thereby giving rise to strong-coupling superconductivity. The emergence of a consistent boundary $x_c$ across the optimal regime in normal and superconducting states provides valuable insights such as the electronic band structure reconstruction and the evolution of the superconducting pairing interaction. These results suggest the universality that superconductivity emerges in close proximity to a strongly correlated phase, a concept that has its origins in copper oxide superconductors (*1*, *3*). The electron correlations in nickelates deserves further investigation, particularly with regard to stripe-like charge order and the spin-wave-like magnetic excitations, potentially contributing to the complex interplay of charge and spin degrees of freedom. The unique characteristics in their strong-coupling superconductivity and possible quantum criticality remains to be investigated (*13*, *42*, *49–52*).

**Materials and Methods**
**Sample preparation**

$La_{1-x}Sr_xNiO_3$ films ($x = 0.04$–$0.28$) were grown on ~5 × 10 mm² $SrTiO_3$ (001) substrates using pulsed-laser deposition. $La_{1-x}Sr_xNi_{1.2}O_y$ polycrystalline targets were ablated by a KrF excimer laser (wavelength 248 nm). $SrTiO_3$ (001) substrates were pre-annealed at 750°C in an oxygen partial pressure of $1 \times 10^{-6}$ Torr to obtain an atomically flat surface. During growth, the substrate temperature was fixed at 520°C and the oxygen partial pressure was 200 mTorr. The laser fluence was 0.8 J/cm² and a repetition rate was 4 Hz. The film thickness was in a range of 6 to 10 nm by adjusting the laser pulse counts. The films were characterized using x-ray diffraction (XRD) techniques with Cu K$a$ source ($\lambda$ = 1.5406 Å). After deposition, the films were placed in a Pyrex glass tube with calcium hydride (CaH$_2$) powder (~0.1 g), wrapped in aluminum foil to prevent direct contact with the powder. The glass tube was sealed under vacuum (below 10 mTorr) using a methane gas torch and a dry pump. The sealed glass tube was then annealed in a tube furnace at 340ºC–370ºC, with ramping and cooling rates of 10ºC/min. The total reduction time was 2.5–4 hours. After the reduction, XRD $\theta$–$2\theta$ symmetric scans were measured to confirm the stabilization of infinite-layer phase. Interface quality was examined scanning tunneling electron microscopy (STEM) for $x$ = 0.16 film.

**Device fabrication for (thermo)electric measurement**

After reduction, a half of the film surface (~2.5 × 10 mm$^2$) was polished to expose SrTiO$_3$ substrates. Photoresist was patterned to make a dumbbell-shaped on-chip thermometer and to protect La$_{1-x}$Sr$_x$NiO$_2$ area, and then ~20 nm-thick Pt/Ti was deposited at room temperature by radio-frequency magnetron sputtering at an Ar gas pressure of 0.5 Pa and a radio-frequency power of 50 W. After photoresist was removed by hot acetone, Pt/Ti thermometers were formed on SrTiO$_3$ substrates.

**Electrical transport measurement**

The measurements of the temperature dependent resistivity $\rho_{xx}$ was measured using a six-point geometry with Au wire bonded on In contact in physical properties measurement system (PPMS, Quantum Design, Inc.). The Hall effect was measured to be linear up to the highest measured magnetic field of 9 T (–9 T).

**Thermoelectric measurement**

The thermoelectric voltages were measured by Keithley 2182A nanovoltmeters. The temperatures were measured by two Pt/Ti thermometers attached on the SrTiO$_3$ substrate (uncovered area with La$_{1-x}$Sr$_x$NiO$_2$) by gold wires using In contact. The Pt/Ti thermometers were calibrated in in-situ zero magnetic field. With silver paste, the sample is bridged on a cupper stage for a heat sink and Bakelite (epoxy resin) plate for a heater stage. A transverse thermal gradient was generated using a resistor attached on the heater stage. This device geometry allows us to measure the Seebeck effect from room temperature down to around 20 K. We confirmed that the thermoelectric voltage was proportional to the temperature gradient and obtained the generated thermoelectric voltage $\Delta V_{xx}$ by subtracting $V_{xx}$ with the heater off from $V_{xx}$ with the heater on. Seebeck coefficients of gold, for instance, are 0.82, 1.02, and 1.34 µV/K at 100 K, 150 K, and 200 K, respectively (*53*). We therefore assume that the contribution of the gold wire is negligible.

**High-magnetic field electrical resistance measurement**

High-field electrical resistance measurements were carried out using a non-destructive pulse magnet at the International MegaGauss Science Laboratory, Institute for Solid State Physics in the University of Tokyo. The maximum field and pulse duration of the pulse magnet were 60 T and 36 ms, respectively. Electrical resistances were measured by the standard 4 probe method, and a 20 kHz AC current of approximately 10 µA was applied to the samples. We also performed complementary measurements of the electrical resistance under low magnetic fields up to 9 T in a physical-property measurement system (Quantum Design, Inc.).

**Theoretical calculations based on density functional theory**

The electronic structure of $La_{1-x}Sr_xNiO_2$ was calculated using Quantum ESPRESSO (*54*), based on the crystal structure of $LaNiO_2$ obtained from neutron powder diffraction measurements (*55*). The cutoff energies for the wave function and the charge density were set to 100 Ry and 400 Ry, respectively. The calculation was performed with a 12 × 12 × 12 ***k*** point grid for sampling the first Brillouin zone. After the DFT calculation, maximally localized Wannier functions were constructed to extract the hopping parameters of the three-orbital tight-binding model. The transport coefficients are then calculated using the Boltzmann equation based on the DFT band dispersion. However, DFT alone does not fully capture the experimental observations, even in the overdoped regime where electron correlations are expected to weaken. To address the discrepancy, we examined the effect of fundamental electron correlations, namely, the mass enhancement and temperature dependence of the scattering rate, for the Ni $3d_{x^2-y^2}$ band (see Supplementary Materials for detailed calculation and analysis). These factors are essential for accurate transport modeling, as the carrier mobility is proportional to the scattering time (or lifetime) and inversely proportional to the effective mass.

**References**

1. P. A. Lee, N. Nagaosa, X. G. Wen, Doping a Mott Insulator: Physics of High-Temperature Superconductivity. *Rev. Mod. Phys.* **78**, 17 (2006).
2. L. Taillefer, Scattering and Pairing in Cuprate Superconductors. *Annu. Rev. Condens. Matter Phys.* **1**, 51–70 (2010).
3. B. Keimer, S. A. Kivelson, M. R. Norman, S. Uchida, J. Zaanen, From Quantum Matter to High-Temperature Superconductivity in Copper Oxides. *Nature* **518**, 179 (2015).
4. J. Bardeen, L. N. Cooper, J. R. Schrieffer, Theory of superconductivity. *Phys. Rev.* **108**, 1175 (1957).
5. D. Li *et al.*, Superconductivity in an infinite-layer nickelate. *Nature* **572**, 624-627 (2019).
6. V. I. Anisimov, D. Bukhvalov, T. M. Rice, Electronic structure of possible nickelate analogs to the cuprates. *Phys. Rev. B* **59**, 7901 (1999).
7. A. S. Botana, M. R. Norman, Similarities and Differences between $LaNiO_2$ and $CaCuO_2$ and Implications for Superconductivity. *Phys. Rev. X* **10**, 011024 (2020).
8. Y. Nomura *et al.*, Formation of a two-dimensional single-component correlated electron system and band engineering in the nickelate superconductor $NdNiO_2$. *Phys. Rev. B* **100**, 205138 (2019).
9. Y. Nomura, R. Arita, Superconductivity in infinite-layer nickelates. *Rep. Prog. Phys.* **85**, 052501 (2022).
10. M. Kitatani *et al.*, Nickelate superconductor—a renaissance of the one-band Hubbard model. *npj Quantum Mater.* **5**, 59 (2020).
11. H. Sakakibara *et al.*, Model Construction and a Possibility of Cupratelike Pairing in a New $d^9$ Nickelate Superconductor $(Nd,Sr)NiO_2$. *Phys. Rev. Lett.* **125**, 077003 (2020).
12. K. Lee *et al.*, Linear-in-temperature resistivity for optimally superconducting $(Nd,Sr)NiO_2$. *Nature* **619**, 288–292 (2023).
13. R. A. Cooper *et al.*, Anomalous Criticality in the Electrical Resistivity of $La_{2-x}Sr_xCuO_4$. *Science* **323**, 603–607 (2009).
14. X. Jiang *et al.*, Interplay between superconductivity and the strange-metal state in FeSe. *Nat. Phys.* **19**, 365–371 (2023).
15. Y.-T. Hsu *et al.*, Transport phase diagram and anomalous metallicity in superconducting infinite-layer nickelates. *Nat. Commun.* **15**, 9863 (2024).
16. M. Hepting *et al.*, Electronic structure of the parent compound of superconducting infinite-layer nickelates. *Nat. Mater.* **19**, 381–385 (2020).
17. H. Lu *et al.*, Magnetic excitations in infinite-layer nickelates. *Science* **373**, 213–216 (2021).
18. B. H. Goodge *et al.*, Doping evolution of the Mott–Hubbard landscape in infinite-layer nickelates. *Proc. Natl. Acad. Sci. U. S. A.* **118**, e2007683118 (2021).
19. M. Rossi *et al.*, A Broken Translational Symmetry State in an Infinite-Layer Nickelate. *Nat. Phys.* **18**, 869 (2022).
20. C. C. Tam *et al.*, Charge density waves in infinite-layer $NdNiO2$ nickelates. *Nat. Mater.* **21**, 1116–1120 (2022).
21. G. Krieger *et al.*, Charge and spin order dichotomy in $NdNiO2$ driven by the capping layer. *Phys. Rev. Lett.* **129**, 027002 (2022).
22. C. T. Parzyck *et al.*, Absence of $3a_0$ charge density wave order in the infinite-layer nickelate $NdNiO_2$. *Nat. Mater.* **23**, 486 (2024).
23. M. Osada *et al.*, Nickelate Superconductivity without Rare-Earth Magnetism: $(La,Sr)NiO_2$. *Adv. Mater.* **33**, 2104083 (2021).

24. M. Osada, K. Fujiwara, T. Nojima, A. Tsukazaki, Improvement of superconducting properties in La$_{1-x}$Sr$_x$NiO$_2$ thin films by tuning topochemical reduction temperature. *Phys. Rev. Mater.* **7**, L051801 (2023).
25. L. E. Chow *et al.*, Pauli-limit violation in lanthanide infinite-layer nickelate superconductors. Preprint at https://arXiv:2204.12606 (2022).
26. W. Wei *et al.*, Large upper critical fields and dimensionality crossover of superconductivity in the infinite-layer nickelate La$_{0.8}$Sr$_{0.2}$NiO$_2$. *Phys. Rev. B* **107**, L220503 (2023).
27. W. Sun *et al.*, Evidence for Anisotropic Superconductivity Beyond Pauli Limit in Infinite-Layer Lanthanum Nickelates. *Adv. Mater.* **35**, 2303400 (2023).
28. Y.-T. Hsu *et al.*, Insulator-to-metal crossover near the edge of the superconducting dome in Nd$_{1-x}$Sr$_x$NiO$_2$. *Phys. Rev. Research* **3**, L042015 (2021).
29. B. Y. Wang *et al.*, Effects of rare-earth magnetism on the superconducting upper critical field in infinite-layer nickelates. *Sci. Adv.* **9**, eadf6655 (2023).
30. D. Li *et al.*, Probing Quantum Confinement and Electronic Structure at Polar Oxide Interfaces. *Adv. Sci.* **5**, 1800242 (2018).
31. R. Joynt, L. Taillefer, The superconducting phases of UPt$_3$. *Rev. Mod. Phys.* **74**, 235–294 (2002).
32. A. Devarakonda *et al.*, Clean 2D superconductivity in a bulk van der Waals superlattice. *Science* **370**, 231–236 (2020).
33. K. Deguchi, M. A. Tanatar, Z. Mao, T. Ishiguro, Y. Maeno, Superconducting Double Transition and the Upper Critical Field Limit of Sr$_2$RuO$_4$ in Parallel Magnetic Fields. *J. Phys. Soc. Jpn.* **71**, 2839–2842 (2002).
34. F. Weickert, P. Gegenwart, In-plane angular dependence of the upper critical field in CeCoIn$_5$. *Phys. Rev. B* **74**, 134511 (2006).
35. Y. Saito *et al.*, Superconductivity protected by spin–valley locking in ion-gated MoS$_2$. *Nat. Phys.* **12**, 144–149 (2016).
36. S. I. Vedeneev, B. A. Piot, A. V. Sadakov, Temperature dependence of the upper critical field of FeSe single crystals. *Phys. Rev. B* **87**, 134512 (2013).
37. S. J. Williamson, Bulk Upper Critical Field of Clean Type-II Superconductors: V and Nb. *Phys. Rev. B*, **2**, 3545–3556 (1970).
38. N. Kurita *et al.*, Determination of the Upper Critical Field of a Single Crystal LiFeAs: The Magnetic Torque Study up to 35 Tesla. *J. Phys. Soc. Jpn.* **80**, 013706. (2011).
39. M. Zehetmayer *et al.*, Mixed-state properties of superconducting MgB$_2$ single crystals. *Phys. Rev. B* **66**, 052505 (2002).
40. D. Nakamura, T. Adachi, K. Omori, Y. Koike, S. Takeyama, Pauli-limit upper critical field of high-temperature superconductor La$_{1.84}$Sr$_{0.16}$CuO$_4$. *Sci. Rep.* **9**, 16949 (2019).
41. T. Sekitani, Y. H. Matsuda, N. Miura, Measurement of the upper critical field of optimally-doped YBa$_2$Cu$_3$O$_{7-\delta}$ in megagauss magnetic fields. *New Journal of Physics* **9**, 47, (2007).
42. K. Mukasa *et al.*, Enhanced Superconducting Pairing Strength near a Pure Nematic Quantum Critical Point. *Phys. Rev. X* **13**, 011032 (2023).
43. J. M. Park *et al.*, Tunable strongly coupled superconductiviity in magic-angle twisted trilayer graphene. *Nature* **590**, 249 (2021).
44. S. Imajo *et al.*, Extraordinary $\pi$-electron superconductivity emerging from a quantum spin liquid. *Phys. Rev. Research* **3**, 033026 (2021).
45. Q. Zhao *et al.*, Isotropic Quantum Griffiths Singularity in Nd$_{0.8}$Sr$_{0.2}$NiO$_2$ Infinite-Layer Superconducting Thin Films. *Phys. Rev. Lett.* **133**, 036003 (2024).

46. F. Laliberté *et al.*, Fermi-surface reconstruction by stripe order in cuprate superconductors. *Nat. Commun.* **2**, 432 (2011).
47. A. Gourgout *et al.*, Seebeck Coefficient in a Cuprate Superconductor: Particle-Hole Asymmetry in the Strange Metal Phase and Fermi Surface Transformation in the Pseudogap phase. *Phys. Rev. X* **12**, 011037 (2022).
48. W. O. Wang, J. K. Ding, E. W. Huang, B. Moritz, T. P. Devereaux, Quantitative assessment of the universal thermopower in the Hubbard model. *Nat. Commun.* **14**, 7064 (2023).
49. P. Gegenwart, Q. Si, F. Steglich, Quantum criticality in heavy-fermion metals. *Nat. Phys.* **4**, 186–197 (2008).
50. Y. Mizukami *et al.*, Extremely strong-coupling superconductivity in artificial two-dimensional Kondo. *Nat. Phys.* **7**, 849–853 (2011).
51. B. Michon *et al.*, Thermodynamic signatures of quantum criticality in cuprate superconductors. *Nature* **567**, 210-222 (2019).
52. K. Wakamatsu *et al.*, Thermoelectric signature of quantum critical phase in a doped spin-liquid candidate. *Nat. Commun.* **14**, 3679 (2023).
53. N. Cusack, P. Kendall, The Absolute Scale of Thermoelectric Power at High Temperature. *Proc. Phys. Soc.* **72**, 898 (1958).
54. P. Giannozzi *et al.*, Quantum espresso: a modular and open-source software project for quantum simulations of materials. *J. Phys. Condens. Matter.* **21**, 395502 (2009).
55. M. A. Hayward, M. A. Green, M. J. Rosseinsky, J. Sloan, Sodium hydride as a powerful reducing agent for topotactic oxide deintercalation: synthesis and characterization of the nickel(I) oxide $LaNiO_2$. *J. Am. Chem. Soc.* **121**, 8843–8854 (1999).
56. M. J. van Setten *et al.*, The pseudodojo: Training and grading a 85 element optimized norm-conserving pseudopotential table. *Comput. Phys. Commun.* **226**, 39–54, (2018).
57. J. P. Perdew, K. Burke, M. Ernzerhof, Generalized gradient approximation made simple. *Phys. Rev. Lett.* **77**, 3865–3868 (1996).
58. L. Bellaiche, D. Vanderbilt, Virtual crystal approximation revisited: Application to dielectric and piezoelectric properties of perovskites. *Phys. Rev. B* **61**, 7877, (2000).
59. G. Pizzi *et al.*, Wannier90 as a community code: new features and applications. *J. Phys. Condens. Matter.* **32**, 165902 (2020).
60. K. Momma, F. Izumi, Vesta 3 for three-dimensional visualization of crystal, volumetric and morphology data. *J. Appl. Crystallogr.* **44**, 1272–1276, (2011).
61. K. Kuroki, R. Arita, "Pudding mold" band drives large thermopower in $Na_xCoO_2$. *J. Phys. Soc. Jpn.* **76**, 083707–083707 (2007).
62. J. R. Yates, X. Wang, D. Vanderbilt, I. Souza, Spectral and fermi surface properties from wannier interpolation. *Phys. Rev. B* **75**, 195121 (2007).
63. G. Grissonnanche *et al.*, Linear-in temperature resistivity from an isotropic planckian scattering rate. *Nature* **595**, 667–672 (2021).

**Acknowledgments**

We thank K. Kisu and S. Orimo for technical support. STEM observations were made with the cooperation of Y. Kodama and T. Konno of Analytical Research Core for Advanced Materials, Institute for Materials Research, Tohoku University. High-field measurements were performed using facilities of the Institute for Solid State Physics, The University of Tokyo.

**Competing interests:**

Authors declare that they have no competing interests.

**Data and materials availability:**

All data are available in the main text or the supplementary materials.

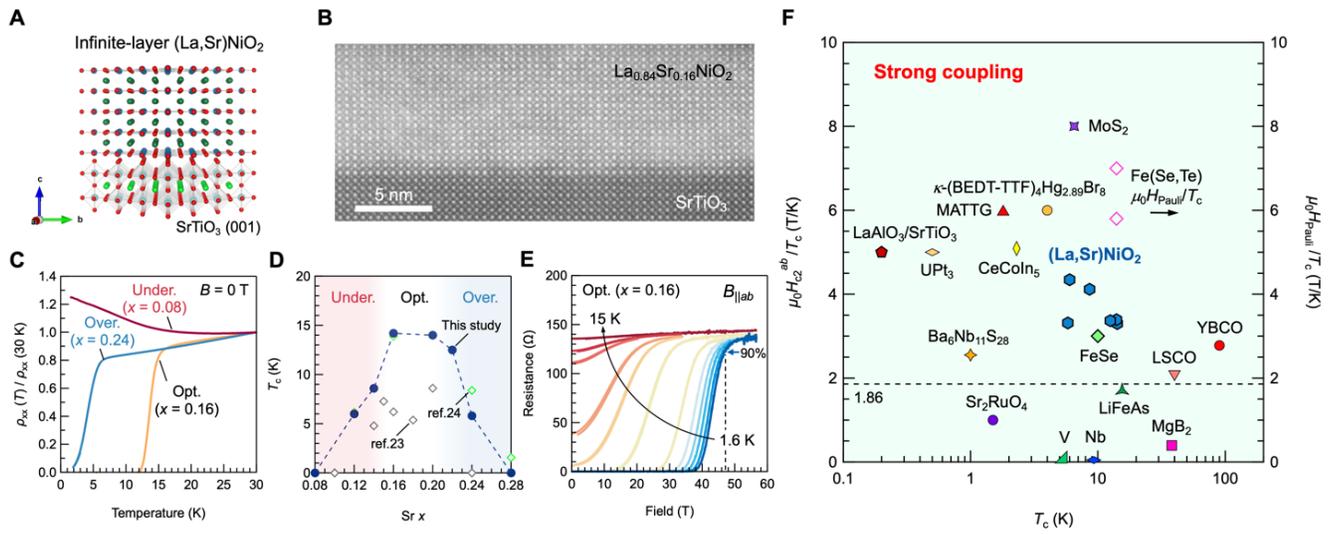

**Fig. 1. Basic properties of superconducting infinite-layer La$_{1-x}$Sr$_x$NiO$_2$ and comparison of pairing strength with other superconductors.** (**A**) Crystal structure of infinite-layer nickelates La$_{1-x}$Sr$_x$NiO$_2$ on SrTiO$_3$. (**B**) HAADF-STEM cross-sectional image of a La$_{0.84}$Sr$_{0.16}$NiO$_2$ thin film grown on SrTiO$_3$ taken along the [100] axis (scale bar 5 nm). (**C**) Normalized resistivities of La$_{1-x}$Sr$_x$NiO$_2$ films with Sr $x$ = 0.08, 0.16, and 0.24 under zero-magnetic field. (**D**) The onset $T_c$ as a function of $x$: Superconducting dome of La$_{1-x}$Sr$_x$NiO$_2$ thin films (Filled blue circles are in this study; gray open diamonds are La$_{1-x}$Sr$_x$NiO$_2$ data from (ref. *23*); green open diamonds are La$_{1-x}$Sr$_x$NiO$_2$ data from (ref. *24*)). (**E**) Resistance of La$_{0.84}$Sr$_{0.16}$NiO$_2$ as a function of magnetic fields parallel to the *ab*-plane at various temperature. (**F**) Survey of superconducting materials (ref. *30–44*) characterized by the upper critical field parallel to the conducting planes $H_{c2}^{ab}$ (Pauli-limited field, $H_{Pauli}$) to superconducting critical temperature $T_c$. The broken line indicates the weak-coupling BCS superconducting limit about 1.86 T/K.

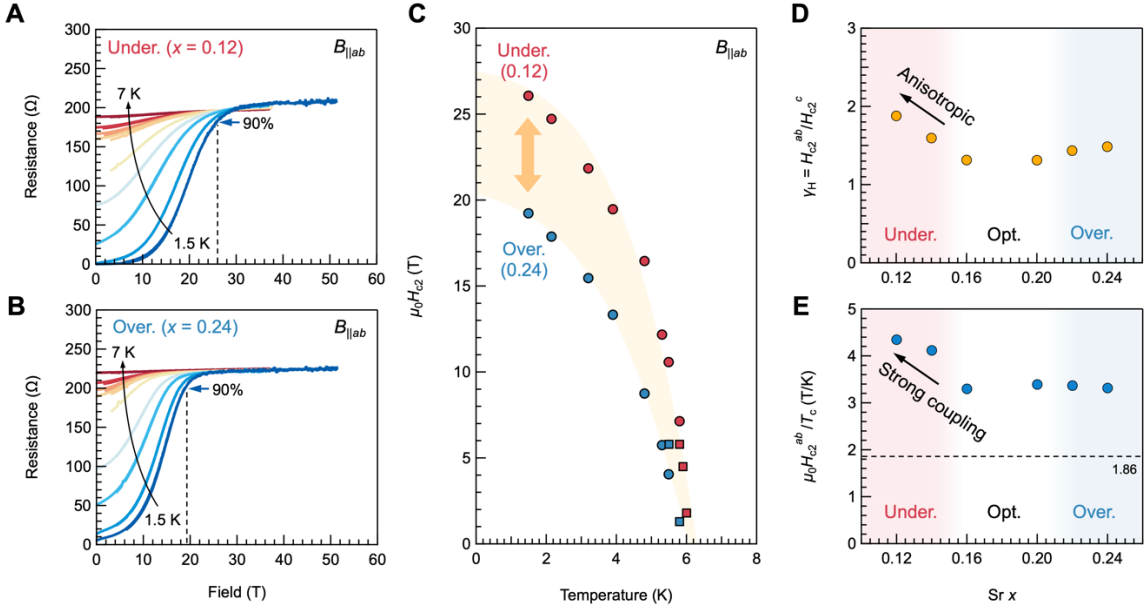

**Fig. 2. Larger upper critical field of underdoped nickelates than that of overdoped, detected by high-field measurements.** Resistance as a function of magnetic fields parallel to the *ab*-plane at various temperature for (**A**) underdoped (Sr $x$ = 0.12) and (**B**) overdoped nickelates (Sr $x$ = 0.24). (**C**) Upper critical field $\mu_0 H_{c2}$ extracted using the 90% normal resistance criterion for underdoped (Sr $x$ = 0.12, red circles) and overdoped nickelates (Sr $x$ = 0.24, blue circles). Orange area is a guide to the eye, indicating difference between underdoped and overdoped nickelates. High- and low-field data were measured under pulsed high-magnetic fields (circles) and continuous magnetic fields (squares), respectively. (**D**) Superconducting anisotropy, $\gamma_H = H_{c2}^{ab}/H_{c2}^{c}$, and (**E**), pairing strength, $\mu_0 H_{c2}^{ab}/T_c$, as a function of $x$ at the lowest measurement temperature of 1.4–1.6 K. The broken line indicates the weak-coupling BCS superconducting limit about 1.86 T/K.

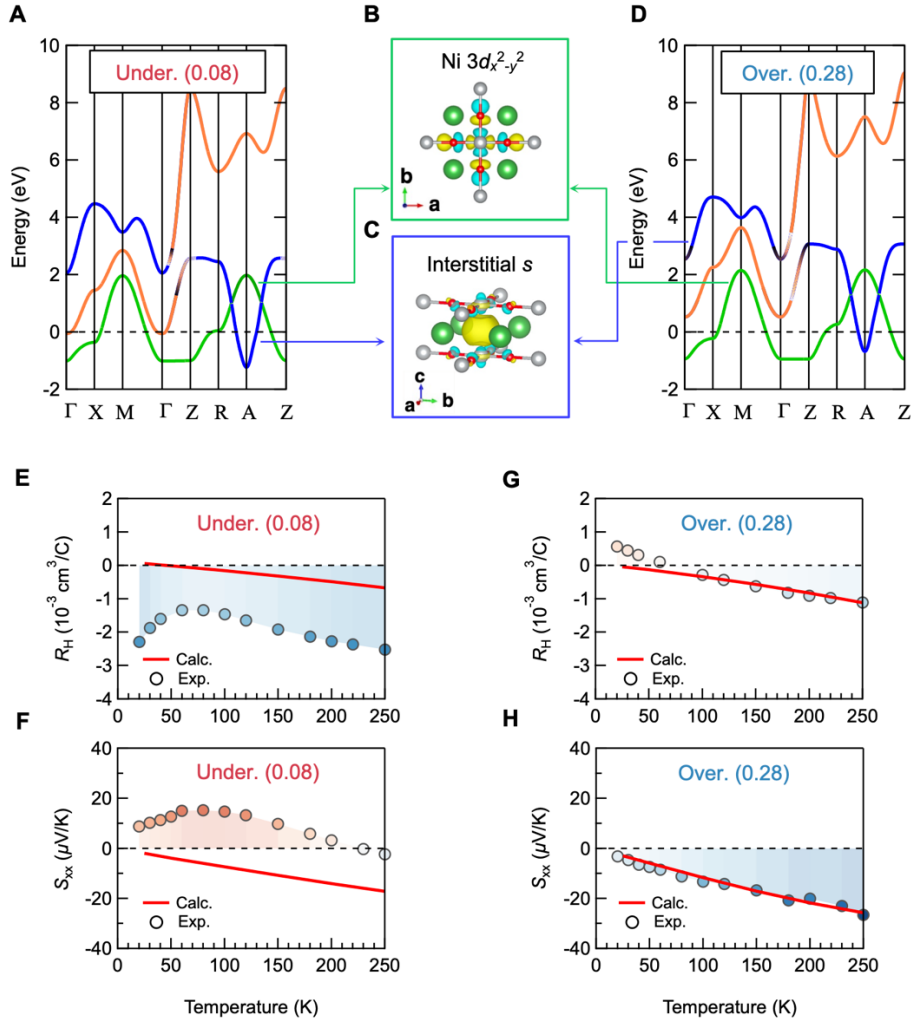

**Fig. 3. Inconsistency between DFT-based calculations and experimental results for Hall and Seebeck coefficients of underdoped nickelates.** (**A**) DFT result for the electronic band structure for underdoped $x = 0.08$. (**B** and **C**) The isosurfaces of contributing Wannier functions of the Ni $3d_{x2-y2}$ orbital and interstitial $s$ orbital, respectively. (**D**) DFT result for the electronic band structure for overdoped $x = 0.28$. (**E** and **F**) Hall coefficients and Seebeck coefficients as a function of temperature of underdoped $La_{0.92}Sr_{0.08}NiO_2$. DFT-based calculations (red line) and experimental results (circles) are depicted. (**G** and **H**) Hall coefficients and Seebeck coefficients as a function of temperature of overdoped $La_{0.72}Sr_{0.28}NiO_2$. DFT-based calculations (red line) and experimental results (circles) are depicted.

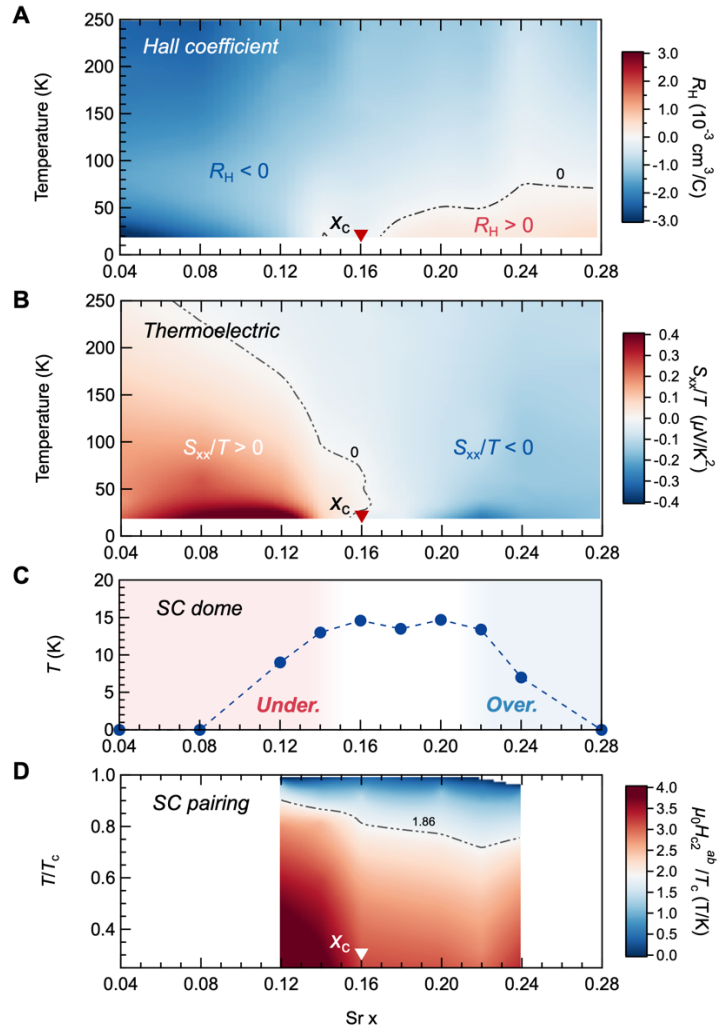

**Fig. 4. Electron correlation-driven sign reversal in Seebeck coefficients for underdoped nickelates, coinciding with the emergence of strong-coupling superconductivity at $x < 0.16$.** Color maps of (**A**) $R_H$ and (**B**) $S_{xx}/T$ at normal conducting state. The broken curves indicate zero. (**C**) Superconducting transition temperatures of $La_{1-x}Sr_xNiO_2$. (**D**) Color map of $\mu_0 H_{c2}^{ab}/T_c$ at superconducting state. The broken curve indicates the weak-coupling BCS superconducting limit about 1.86 T/K. The boundary doping concentrations $x_c$ are shown as triangles.